\documentclass[twocolumn]{openjournalold}

\usepackage{orcidlink}
\usepackage[T1]{fontenc}
\usepackage{ae,aecompl}
\usepackage{graphicx}	
\usepackage{amsmath}	
\usepackage{amssymb}	

\usepackage{natbib}
 
\usepackage{hyperref}
\hypersetup{
	colorlinks=true,
	urlcolor=blue,    
	linkcolor=black,  
	citecolor=blue,   
}

\defcitealias{Pittordis_2018}{PS18}
\defcitealias{Pittordis_2019}{PS19}
\defcitealias{Badry_2021}{ERH}
\defcitealias{Banik_2018_Centauri}{BZ18}

\hypersetup{citecolor=blue, 
            linkcolor=red, 
            menucolor=blue, 
            urlcolor=blue}  

\begin{document}

\title{The distribution of misalignment angles in multipolar planetary nebulae}
\date{December 2024}

\author{Ido Avitan\,\orcidlink{0009-0000-8094-0614}}
\affiliation{Department of Physics, Technion - Israel Institute of Technology, Haifa, 3200003, Israel; ido.avitan@campus.technion.ac.il; soker@physics.technion.ac.il}

\author{Noam Soker\,\orcidlink{0000-0003-0375-8987}} 
\affiliation{Department of Physics, Technion - Israel Institute of Technology, Haifa, 3200003, Israel; ido.avitan@campus.technion.ac.il; soker@physics.technion.ac.il}

\begin{abstract}
We measure the projected angle on the plane of the sky between adjacent symmetry axes of tens of multipolar planetary nebulae and find that the distribution of these misalignment angles implies a random three-dimensional angle distribution limited to $\lesssim 60^\circ$. We identify a symmetry axis as a line connecting two opposite lobes (bubbles) or clumps. We build a cumulative distribution function of the projected angles $\alpha$ and find that an entirely random distribution of the three-dimensional angles $\delta$ between adjacent symmetry axes, namely, uncorrelated directions, does not fit the observed one. A good fit to the observed distribution is a limited random distribution of the three-dimensional angle between adjacent symmetry axes, i.e., random distribution in the range of $20^\circ \lesssim \delta \lesssim 60^\circ$. We assume that a pair of jets along the angular momentum axis of an accretion disk around the companion shape each symmetry axis. The limited random distribution might result from two sources of angular momentum to the accretion disks with comparable magnitude: one with a fixed direction and one with a stochastic direction variation. We discuss a scenario where the fixed-axis angular momentum source is the binary orbital angular momentum, while the stochastic source of angular momentum is due to the vigorous envelope convection of the mass-losing giant progenitor. 
\end{abstract}  


\section{Introduction}
\label{sec:Introduction}

Most planetary nebulae (PNe) have one or more pairs of structural features on opposite sides of the center. The structural features include bubbles (a bubble is a faint zone closed by a bright rim), lobes (a lobe is a faint zone with a partial outer bright rim open to the far side), clumps (also called ansae), nozzles (a nozzle is a narrow opening in the PN shell), ears (an ear is a protrusion with a base smaller than the PN main shell and a cross-section that decreases outward), or arcs (rims).
Hereafter, we refer to the two opposite structural features as poles and the line connecting them as a symmetry axis. 
PNe that have two or more symmetry axes are termed multipolar PNe (e.g., \citealt{SahaiTrauger1998, Steffenetal2013, Huangetal2016, RechyGarciaetal2020, Bandyopadhyayetal2023, Wenetal2023}. Multipolar PNe with two symmetry axes are called quadrupole PNe (e.g., \citealt{Manchadoetal1996b, Guerreroetal2013}).  

Multipolar PNe have been the center of some past studies (see the recent review by \citealt{Kwok2024}), e.g., \cite{Velazquez2012} modeled precessing jets with varying velocity as the cause of multipolar PNe. \cite{Chongetal2012} suggested that multipolar structures are common in PNe, much more common than what is directly observed (without analysis). In this study, we will study only the images of multipolar PNe, referring only to their appearance on the plane of the sky. 

Most studies attribute each symmetry axis to a pair of jets launched by a binary system progenitor (e.g.,  \citealt{Morris1987, Soker1990AJ, SahaiTrauger1998, AkashiSoker2018,   EstrellaTrujilloetal2019, Tafoyaetal2019, Balicketal2020, RechyGarciaetal2020, Clairmontetal2022, Danehkar2022, MoragaBaezetal2023, Ablimit2024, Derlopaetal2024, Mirandaetal2024, Sahaietal2024} for a list of a small fraction of all papers;  
\citealt{Baanetal2021} present an alternative scenario based on fallback material). In most cases, the binary system progenitor experiences a common envelope evolution of a main sequence star orbiting inside the envelope of an asymptotic giant branch (AGB) star; a red giant branch (RGB) star as the primary star is also possible (e.g., \citealt{Hillwigetal2017, Sahaietal2017, Jonesetal2020, Jonesetal2022, Jonesetal2023}). 

The pair of jets might change direction during the PN formation phase. One pattern is the precession of the jet axis, as observed in some PNe (e.g., \citealt{Guerreroetal1998, Mirandaetal1998, Sahaietal2005, Boffinetal2012, Sowickaetal2017, RechyGarciaetal2019, Guerreoetal2021, Clairmontetal2022}).  
Another pattern is jet-launching episodes with non-monotonic varying axis directions. \cite{Soker2025Bright} suggests that most of the brightest PNe are multipolar formed in a violent process with non-continuous direction variation of the jets' axis. We return to this point in our discussion in Section \ref{sec:Summary}.  

In this study, we examine multipolar PNe, where different pairs of lobes indicate separate jet-launching episodes rather than precession or another continuous change of direction. In Section \ref{sec:Sample}, we present the sample of multipolar PNe with the projected angles between their symmetry axes. In Section \ref{sec:Angles} we analyze the angle distribution and summarize our results in Section \ref{sec:Summary}

\section{The sample of multipolar PNe}
\label{sec:Sample}

In Tables \ref{Tab:Table1} and \ref{Tab:Table2}, we list the 40 PNe and 5 pre-PNe, respectively, for which we could find two or more well-defined symmetry axes. In the second column, we give the angle between the axes in each PN (in degrees), referencing an image in the last column. 
The sample includes all multipolar PNe and pre-PNe we could find in the literature and from private communications (e.g., the acknowledgment section and searching the Astrophysics Data System-ADS), for which the images were clear enough to define two or more axes. 
\begin{table*}
  \caption{Planetary nebulae with binaries and jets}
\begin{minipage}{0.5\textwidth}  
    \begin{tabular}{| p{1.5cm} | p{1.5cm} | p{0.8cm}|  }
\hline  
PN & $\alpha$  & Ref \\
\hline  
M 1-16 & 6 & GM23 \\
M 2-46   &  8 & Ma96 \\ 
Hen 2-96 & 8 & We24 \\
M 1-61 & 12; 55 & Sa11\\
J320      & 15; 18 & Ha04 \\
Pe 1-1 &15; 23; 48 & We23 \\ 
Hen 2-73 & 16 & We24 \\
Hen 2-158 & 17; 47 & We23 \\
NGC 6790  & 18 & Hs14 \\
NGC 6886 & 18 & PNIC \\
NGC 6072 & 19; 38 & Kw10 \\
NGC 6644 & 20 & Sc92 \\
He 2-47 & 21; 22; 46 & Sa00 \\
Hen 2-447 & 24 & PNIC \\ 
M 1-75 & 25 & Ma96 \\ 
M1-37     & 27; 36 & Sa00 \\
Me 2-2 & 27; 48 & Sa98 \\
IC 5117 & 29 & Hs14 \\
NGC 6309  & 30 & Ru15 \\
NGC 5307 & 31; 46 & PNIC \\
\hline  
\end{tabular}
\end{minipage} \hfill
\begin{minipage}{0.5\textwidth}  
    \begin{tabular}{| p{1.5cm} | p{1.5cm} | p{0.8cm}|  }
\hline  
PN & $\alpha$  & Ref \\
\hline  
NGC 6302 & 32 & Sc92 \\
IC 4634 & 33 & Gu08 \\
NGC 6572 & 33 & Ak16 \\
M 1-31 & 40; 42 & PNIC\\
Kn 26     & 41    & Gu13 \\
IC 4846 & 42 & Mi01 \\
NGC 7026 & 46 & Cl13 \\
KjPn 8 & 49 & Lo00 \\
M 4-14 & 51 & Ma96 \\
M 1-33 & 52 & PNIC \\
NGC 6445 & 56 & Sc92 \\
NGC 2440 & 58 & Lo98\\
M 3-28 & 59 & Ma96 \\
NGC 7027 & 61 & Mo23 \\
Hen 2-115 & 61 & Sa98 \\
NGC 2371 & 63 & GG20 \\
M 3-35    & 72 & PNIC \\ 
K 3-24 & 74 & Ma96 \\
NGC 5315 & 80 & PNIC \\
M 1-59    & 82     & Go24 \\
\hline  
\end{tabular}
     \end{minipage} \hfill
  \label{Tab:Table1}\\
\small 
Note:  Projected angle between jet axes in multipolar PNe (second column in degrees). 
\newline
Abbreviation: PN: planetary nebula; Ref: image reference. 
\newline
References:
Ak16: \cite{AkrasGonclves2016} and \cite{Bandyopadhyayetal2023};
Cl13: \cite{Clarketal2013}; 
GG20: \cite{GomezGonzalezetal2020};
GM23: \cite{GomezMunozetal2023};
Go24: \cite{Goldetal2024};
Gu08: \cite{Guerrero2008}; 
Gu13: \cite{Guerreroetal2013}; 
Ha04: \cite{Harmanetal2004};
Hs14: \cite{Hsiaetal2014};
Kw10: \cite{Kwoketal2010};
Lo98: \cite{Lopezetal1998};
Lo00: \cite{Lopezetal2000};
Ma96: \cite{Manchadoetal1996}; 
Mi01: \cite{Mirandaetal2001};
Mo23: \cite{MoragaBaezetal2023};
PNIC: HST archive images from the Planetary Nebula Image Catalogue (PNIC) of Bruce Balick \citep{Balick2006};
Ru14: \cite{Rubioetal2015};
Sa98: \cite{SahaiTrauger1998};
Sa00: \cite{Sahai2000};
Sa11: \cite{Sahaietal20112011};
Sc92: \cite{Schwarzetal1992};
We23: \cite{Wenetal2023};
We24: \cite{Wenetal2024}. 
%
 %
\end{table*}
\begin{table}
  \caption{Misalignment angles in multipolar pre-PNe}
 \begin{center}
    \begin{tabular}{| p{2.4cm} | p{1.0cm} | p{0.8cm}|  }
\hline  
PN & $\alpha$  & Ref \\
\hline  
IRAS04395+3601   &  15 & Tr02 \\ 
IRAS16594-4656 & 17; 23 & Hr99 \\ 
IRAS19024+0044 & 32 & Sa05 \\
IRAS17047-5650 & 47 & Sa07a \\ 
IRAS19475+3119 & 62 & Sa07b \\
\hline  
\end{tabular}
  \label{Tab:Table2}\\
\end{center}
\small 
Note:  Similar to Table \ref{Tab:Table1} but for proto-PNE. 
\newline
References: Hr99: \cite{Hrivnaketal1999}; Sa05: \cite{Sahaietal2005Starfish}; Sa07a: \cite{Sahaietal2007a}; Sa07b: \cite{Sahaietal2007b}; Tr02: \cite{Trammelletal2002}; 
%
 %
\end{table}

\begin{figure}[h]
\begin{center}
\includegraphics[trim=0cm 15.0cm 5.5cm 0cm,scale=0.59]{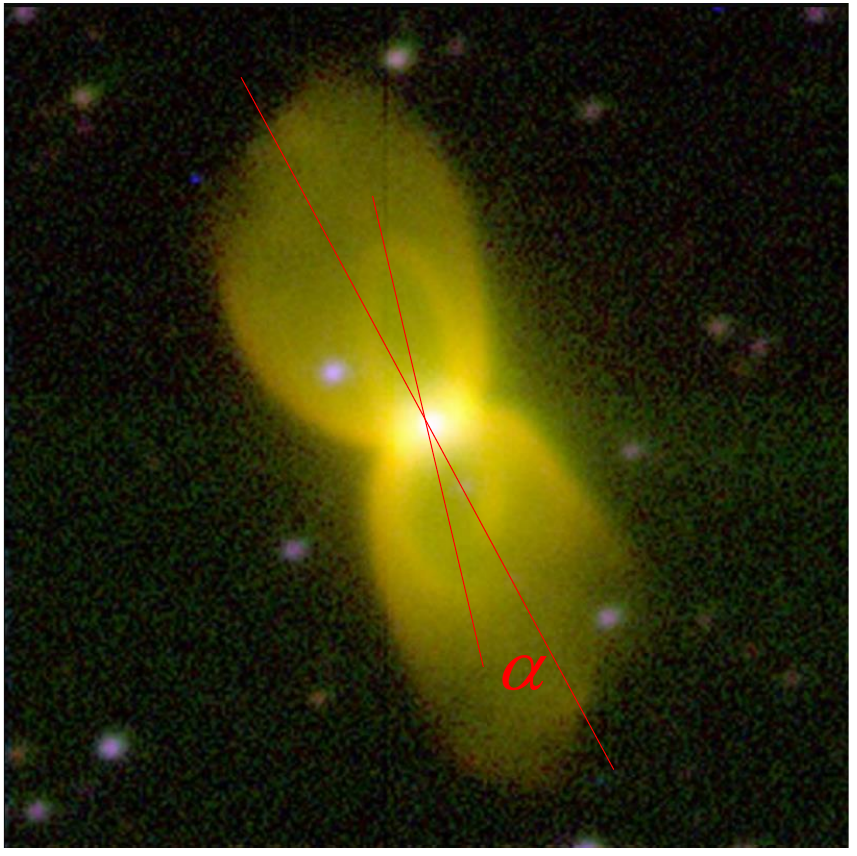} 
\caption{An image of the multipolar (Quadrupolar)  PN M~2-46 adapted from \cite{Manchadoetal1996} with our marks in red. This is an example of one pair of lobes entirely inside the other. 
}
\label{Fig:M246}
\end{center}
\end{figure}

We present several specific examples. The first is PN M 2-46. We present its image in Figure \ref{Fig:M246}, adapted from \cite{Manchadoetal1996}, as an example of a case with one pair of lobes entirely inside the other pair. 

\begin{figure}
\begin{center}
\includegraphics[trim=0cm 23.0cm 5.5cm 0cm,scale=0.62]{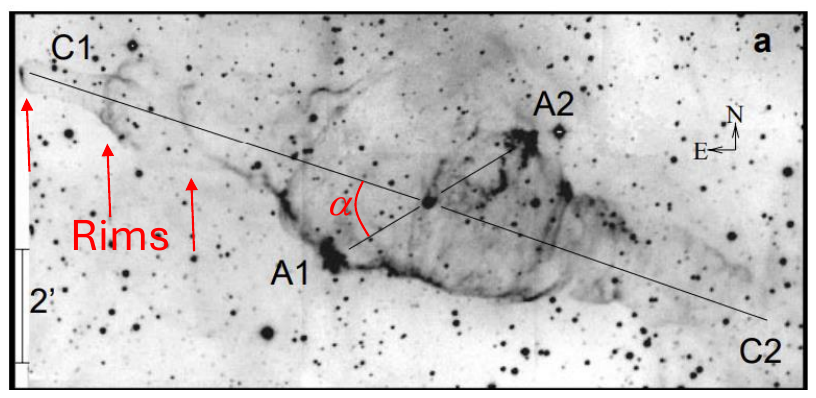} 
\caption{An image of the multipolar PN KjPn~8 adapted from \cite{Lopezetal2000} with their marks in black and our marks in red. The three rims suggest three jet-launching episodes that inflated the large lobes. In this study, we consider them to be one event.  
}
\label{Fig:KjPn8}
\end{center}
\end{figure}

In Figure \ref{Fig:KjPn8} we present an image of the PN KjPn~8 adapted from \cite{Lopezetal2000}. They already marked the two symmetry axes (black lines); we mark the projected angle between the two axes $\alpha$. In the PN KjPn, the small pairs of lobes (A1-A2) are almost entirely inside the large pairs of lobes (C1-C2). In addition, we mark three rims in the eastern lobe with red arrows. Three jets probably compressed these three rims; we do not find evidence of three jet-launching episodes in the western lobe. These three rims without counterparts in the opposite lobe suggest that our analysis might miss some jet-launching episodes with a slight misalignment angle. We return to this point in Section \ref{sec:Angles}.     

\begin{figure}
\begin{center}
\includegraphics[trim=0cm 18.5cm 5.5cm 0cm,scale=0.72]{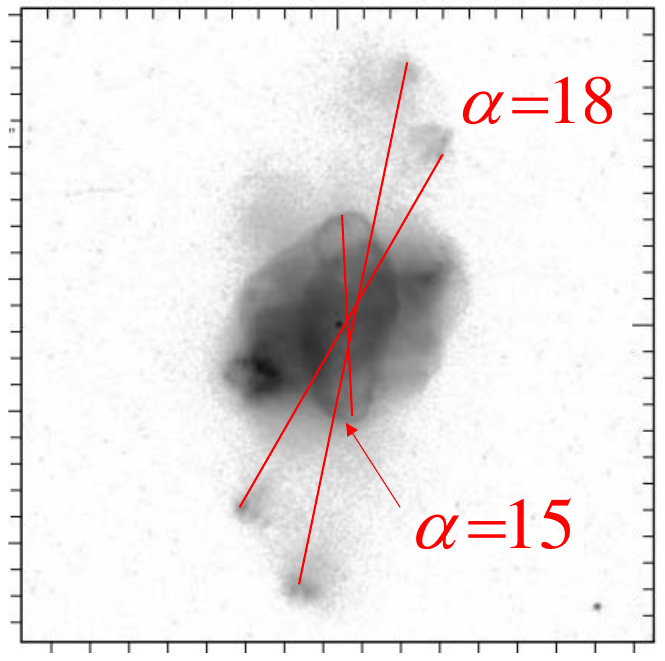} 
\caption{An image of the multipolar J320 adapted from \cite{Harmanetal2004} with  our marks in red. This PN demonstrates uncertainties in the exact poles: the location of the four clumps forming two pairs and the two bubbles of the third pair. The lines we chose by the bright zone of the poles of each axis do not cross exactly at the center. Angles between axes are in degrees. The uncertainties here are about $2^\circ$. However, these do not change our conclusions on the best statistical angle distribution.  
}
\label{Fig:J320}
\end{center}
\end{figure}

Figure \ref{Fig:J320} presents the PN J320, where we identify three symmetry axes, two with pairs of clumps and one with a pair of bubbles. We present this PN to demonstrate a case where the location of the poles is not well defined and where the symmetry axes that we draw by the poles miss the center somewhat. The symmetry lines we draw in our sample can fall in a typical section of about $5 ^\circ$, namely, the typical uncertainties in the projected angles are about $\pm 3^\circ$; in some cases smaller, in some larger. However,  because we study the cumulative distribution function (Section \ref{sec:Angles}), such uncertainties have negligible influence on our conclusions. 

\begin{figure}
\begin{center}
\includegraphics[trim=0cm 12.5cm 5.5cm 0cm,scale=0.50]{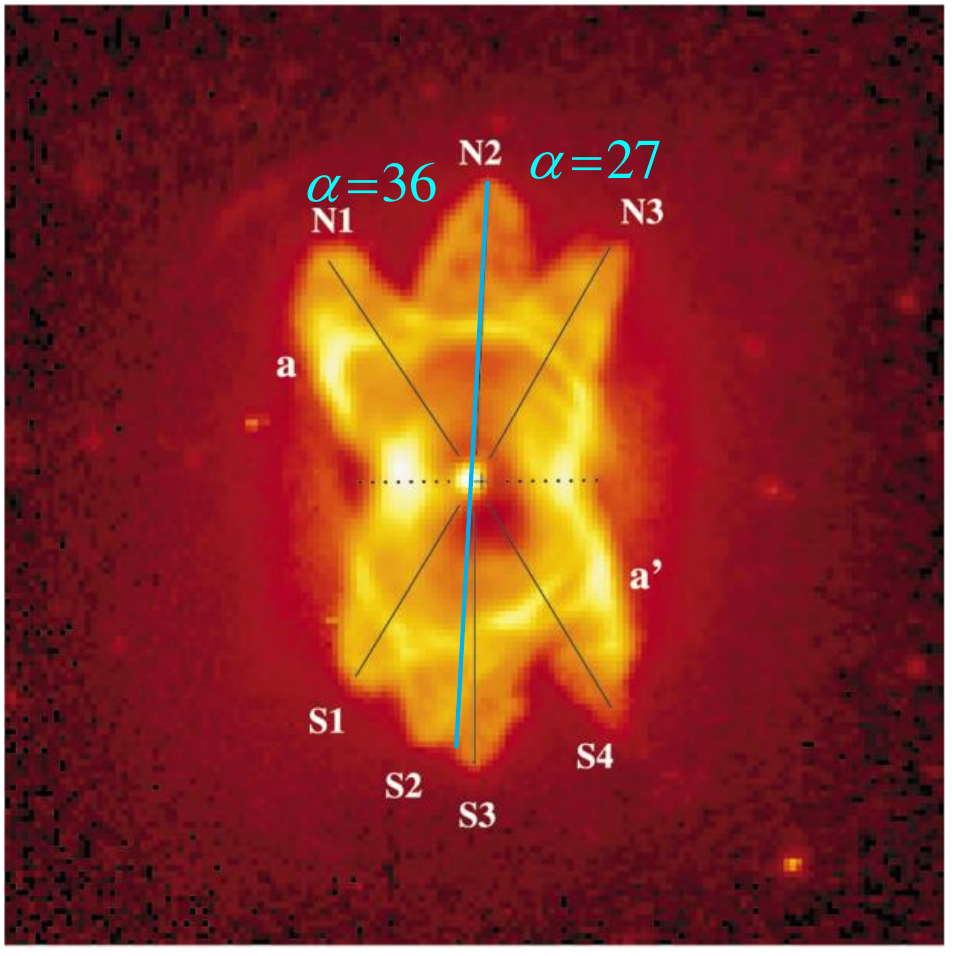} 
\caption{An image of the multipolar M1-37 adapted from \cite{Sahai2000}. The original marks are in black, and ours are in pale blue. \cite{Sahai2000} marks three symmetry axes. He takes the large lobe in the south. We consider the two lobes in the south to be related to the same jets-launching episode and draw a line from the north lobe through the center of the PN. We agree on the other two axes. This again demonstrates 1-2 degrees of uncertainties in the angles.   
}
\label{Fig:M137}
\end{center}
\end{figure}

The PN M1-37 that we present in Figure \ref{Fig:M137} demonstrates a case with two lobes very close to each other, the two in the south. Since there is only one lobe on the other side, the north, we take the two south lobes to belong to the same jet-launching episode. However, we may miss a fourth pair very close to the north-south line that \cite{Sahai2000} drew. We draw a line from the edge of the north lobe through the center; the other side of the line falls between the two neighboring south lobes; for the two other symmetric axes, we take \cite{Sahai2000} marked. This further suggests we might miss small projected angles, $\alpha \lesssim 15^\circ$. 

Some PNe that studies (e.g., \citealt{Guerreroetal2020}) find multipolar do not have clear morphologies for us to measure the angles from the images alone, e.g., IC 4776 \citep{RechyGarciaetal2020}, or are too messy, e.g.,  NGC 5189 \citep{Sabinetal2012} and M 1-26 that might even have three axes, but only one is well defined (image in, e.g., \citealt{RechyGarciaetal2020}). NGC 6210, which is in the list of \cite{Guerreroetal2020} of multipolar PNe, is a messy PN that has a precessing jet pair (e.g., \cite{RechyGarciaetal2020} that we do not consider in this study. \cite{Guillenetal2013} study the multipolar PN NGC 6058. We do not include this PN because we could not identify clear symmetry axes projected on the plane of the sky. Future studies should add more multipolar to our study, those classified by three-dimensional analysis rather than the image on the plane of the sky.   

\section{The distribution of misalignment angles}
\label{sec:Angles}

\begin{figure}[t]
\begin{center}
\includegraphics[trim=3.7cm 1.0cm 5.0cm 1.8cm,scale=0.64]{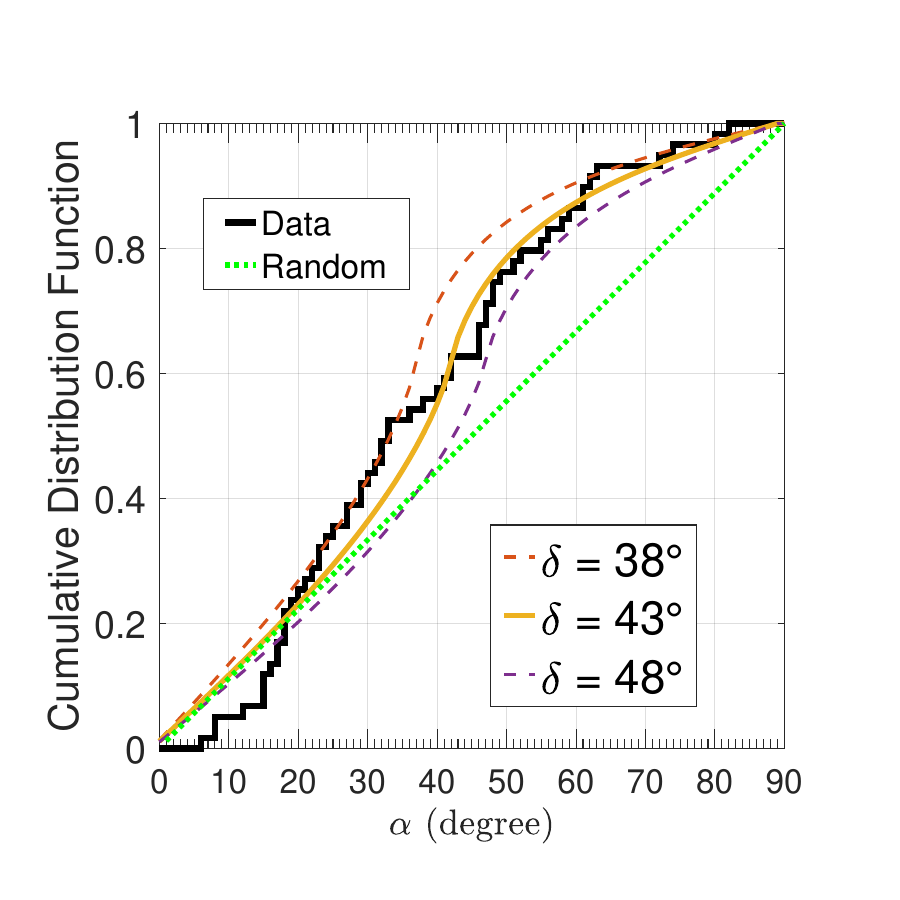} 
\caption{Cumulative distribution functions $W_\alpha$ of projected angels between symmetry axes. The black step function is $W_\alpha$ from the data of 40 PNe and 5 pre-PNe with 58 angles. The green-doted line is $W_\alpha$ for a fully random distribution of angles in their dimension $\delta$; namely, the two or three symmetry axes do not correlate. 
The three other lines are $W_\alpha$ for a constant three-dimensional angle between the two axes, with values given in the inset. The $\delta=43^\circ$ is the best fit for a constant value of $\delta$.  
}
\label{Fig:CDFconstant}
\end{center}
\end{figure}

 We build the cumulative distribution function $W_\alpha (\alpha)$, which is the fraction of objects with a projected angle $\alpha_i$ in the range of $0 \le \alpha_i \le \alpha$. In Figure  \ref{Fig:CDFconstant} we present this function by the thick-black line (the step function). With a green-dotted line, we show a theoretical $W_\alpha$ for the case where the two jet axes of adjacent axes are entirely random, i.e., uncorrelated. A fully random distribution does not fit the observed distribution. We have tried a constant three-dimensional angle between any two axes. The best fit is for $\delta_f=43^\circ$, as shown by the solid-orange line in Figure \ref{Fig:CDFconstant}. We compare two other theoretical cases with $\delta=\delta_f-5^\circ=38^\circ$ and $\delta=\delta_f+5^\circ=48^\circ$. 

We measure the maximum vertical distance $D_{\rm max}$ of each theoretical line from the data. The best fit theoretical line is the one with the minimum value of $D_{\rm max}$. Because we might miss very small projected angles, as the two jet-inflated lobes/ears might merge, we do not include angles of $\alpha \lesssim 15 ^\circ$ in measuring $D_{\rm max}$.
The best fit for a constant value of three-dimensional angle $\delta$, i.e., $\delta=43^\circ$, has $D_{\rm max} =0.115$ at $\alpha=33^\circ$.  

\begin{figure}[t]
\begin{center}
\includegraphics[trim=3.7cm 1.0cm 5.0cm 1.1cm,scale=0.64]{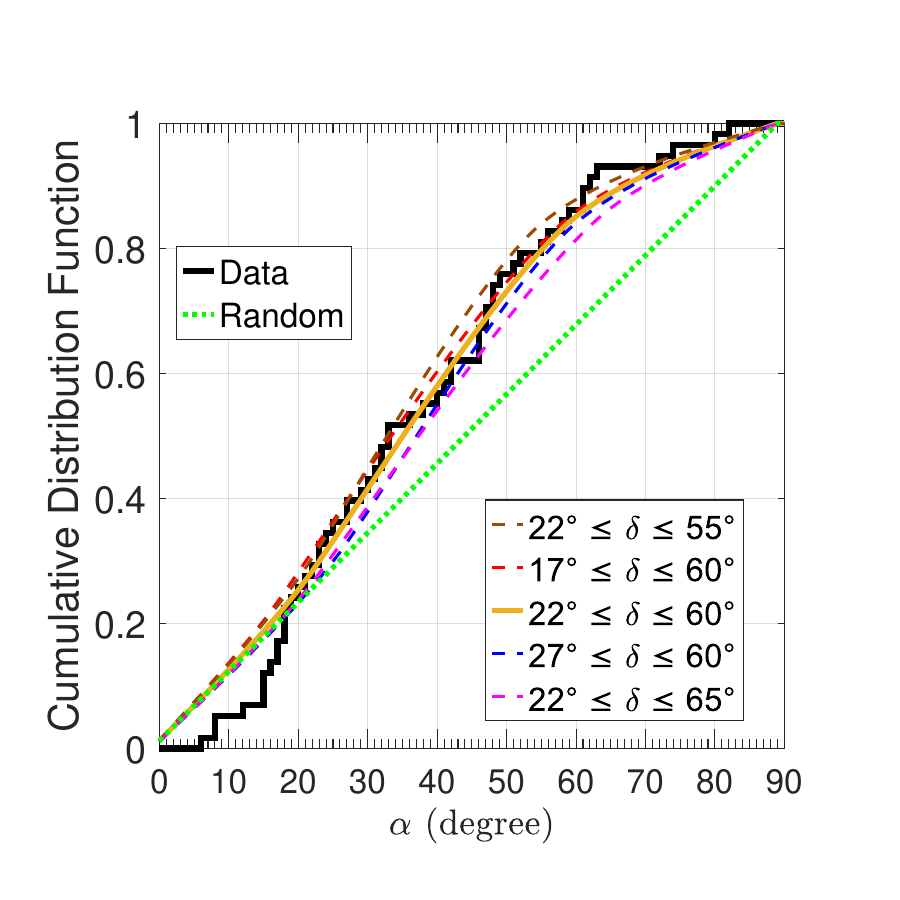} 
\caption{Cumulative distribution functions $W_\alpha$ of projected angels between symmetry axes. The black step function and the green-doted lines are the data and fully random distribution, respectively, as in Figure \ref{Fig:CDFconstant}. The other five lines are for a random distribution of the three-dimensional angle $\delta$ limited by lower and upper bounds, as the inset indicates. The solid-orange line depicts the best fit.   
}
\label{Fig:CDFRandom}
\end{center}
\end{figure}

We next build a theoretical function with a random distribution of three-dimensional angles $\delta$ within a range bounded by a minimum allowed angle $\delta_d$ and a maximum allowed angle $\delta_u$. We find the best fit to the observed distribution to be $\delta_d=22^\circ$, and $\delta_u=60^\circ$, i.e., $\delta$ is entirely random in the range of $22^\circ \lesssim \delta \lesssim 60^\circ$; this is the $W_\alpha(22,60)$ function. Figure \ref{Fig:CDFRandom} presents this case with a solid-orange line, which has $D_{\rm max}= 0.056$ at $\alpha=63^\circ$. For comparison, we present four other cases in Figure  \ref{Fig:CDFRandom}, having larger values of $D_{\rm max}$. The bounded random distribution fits very well with the observed distribution, much better than a constant $\delta$ distribution that we present in Figure \ref{Fig:CDFconstant}. Again, we might miss small angles in observed PNe, and we do not include the range of $\alpha \lesssim 15^\circ$ in determining the best fit.  

One possible explanation for the very good match of the limited random function $W_{\rm \alpha}(22,60)$ with the data and the mismatch of the fully random function (green dotted line) with the data is the presence of two sources of angular momentum with comparable magnitude. An accretion disk launches two opposite jets along its angular momentum axis.  Consider one source of angular momentum with a constant direction and one completely or almost completely random source. If they are of comparable magnitude, or the random is somewhat smaller than the fixed angular momentum component, the two angular momenta add to a new angular momentum direction within a few tens of degrees from the fixed angular momentum axis.
The angles between the axes of two consecutive jet-launching episodes will have a random component within a range smaller than $90^\circ$.  

\section{Summary and discussion}
\label{sec:Summary}

We examined images of tens of PNe and pre-PNe. We measured the projected angle on the plane of the sky, $\alpha$, between adjacent symmetry axes that researchers attribute to shaping by pairs of opposite jets (Tables \ref{Tab:Table1} and \ref{Tab:Table2}). By the inspection of images without deeper analysis, like velocity measurement, we might miss cases with small projected angles $\alpha$, as some PNe suggest (Figures \ref{Fig:KjPn8} and \ref{Fig:M137}). 
This study does not refer to precessing jets continuously changing direction; we only consider separated symmetry axes. 

We draw the observed cumulative distribution function $W_\alpha$ in Figures \ref{Fig:CDFconstant} and \ref{Fig:CDFRandom} (black step function). We find that an entirely random distribution of the angle $\delta$ between symmetry axes in three dimensions (green-dotted line) does not fit the observed distribution. We plot two other theoretical cumulative distribution functions, one of a constant three-dimensional angle $\delta$ between the two symmetry axes (Figure \ref{Fig:CDFconstant}), and one of a random distribution between two angles (Figure \ref{Fig:CDFRandom}). Not considering small angles of $\alpha \lesssim 15^\circ$ that we might miss in observations, we find the best fit among these functions to be $W_\alpha(22,60)$, a limited random distribution between $22^\circ$ and $60^\circ$.

We suggested (Section \ref{sec:Angles}) that the limited random distribution of three-dimensional angles $\delta$ might result from two sources of angular momentum to the accretion disks that launch the two or more pairs of jets that shaped each multipolar PN. One source has a fixed angular momentum axis, while the other adds an angular momentum with a random direction. The two sources supply angular momenta of comparable magnitude. 

From the 40 PNe in our analysis, only M~2-46 is in the catalog of PNe with central binary systems built and maintained by David Jones \citep{JonesBoffin2017, BoffinJones2019, Jones2025}; see also \cite{ChenPetal2025}. M~2-46 is different from most other multipolar PNe in that one pair is much smaller than the other and inside it (Figure \ref{Fig:M246}).  
 Recently, \cite{ChenPetal2025} find variability with a period of 0.0490 in Kn 26, suggesting it is a close binary system. The two pairs of Kn 26 differ substantially in size and shape, similarly to the two pairs of M~2-46.   

\cite{Soker2023IAUPN} examined the morphologies of the brightest PNe in the Galaxy and concluded that most, but not all, bright PNe tend to be multipolar and possess a minor to medium degree of departure from pure point symmetry. He further argued that violent binary interaction shaped the brightest PNe; triple-star interaction is not common as it leads to messy PNe, which most bright PNe do not. 

We end this study by speculating on a scenario of violent binary interaction that includes a component of angular momentum with a fixed axis and one with a random axis, both of which have similar magnitudes.

Consider an AGB (or an RGB) PN progenitor with vigorous envelope convection and pulsation. We propose that the binary companion that launches the jets has an eccentric orbit and it experiences two to a few periastron passages before it enters a common envelope evolution. We assume that during and following periastron passages, the companion launches the energetic pairs of jets that shape the pairs of lobes/bubbles/ears/clumps we used to define multipolar PNe in this study. Near periastron passages, the companion accretes mass from the closest envelope parts to its orbit. The vigorous convection implies large convective cells protrude the envelope in a stochastic manner. The closest envelope part to the jet-launching companion might not be in the equatorial plane but an uprising convective cell off the equatorial plane. 

Consider a protrusion at a distance $h_c$ from the equatorial plane. A typical value is about half the typical size of a convective cell. In the mixing length theory, the size of a convective cell is about the pressure scale height, $l_P$, or somewhat (up to a factor of about 2) larger. In AGB stars, $l_P \simeq 0.3 R_{\rm G}$ where $R_{\rm G}$ is the AGB radius. Therefore, we take   
$h_c \approx 0.5 l_P \simeq 0.15 R_{\rm G}$.  
The random component due to the convective cells has two terms. The first is due to the orbital motion and the distance of the convective protrusion from the equatorial plane (the orbit) and has a radial (negative or positive) direction  
\begin{equation}
   j_r^{\rm conv} = h_c  v_{\rm orb} \approx 0.15 l_P v_{\rm orb}.
    \label{eq:ConvRAM}
\end{equation}
The second term is due to the radial velocity of the convective cell protrusion, $v_{\rm conv}$; the angular momentum direction is in the negative or positive motion of the companion at periastron passage. The convective speed is the order of the sound speed in the outer envelope zones, which is somewhat below the orbital velocity at the stellar surface.  The magnitude of this term is therefore 
\begin{equation}
   j_\theta^{\rm conv} = h_c  v_{\rm conv} \approx 0.15 l_P v_{\rm conv} \approx 0.1 l_P v_{\rm orb}.  
    \label{eq:ConvTAM}
\end{equation}
 
The fixed direction component is due to the orbital motion. Consider the companion moving at a distance of $\Delta a$ from the surface in the equatorial plane. It might pass even inside the outskirts of the envelope during maximum pulsational expansion or graze the envelope in the grazing envelope evolution. If the companion is inside or grazes the envelope, the density gradient supplies the net accreted angular momentum; it has a scale height of $\simeq l_P$, and since mass accretion occurs from both sides, we should crudely take $\Delta a \approx 0.5 l_P$. The specific angular momentum of the fixed-axis component one
\begin{equation}
   j_z^{\rm orb}= \Delta a ~v_{\rm orb} \approx 0.5 l_p ~v_{\rm orb}, 
    \label{eq:ConstantAM}
\end{equation}
where we take the $z$ axis along the orbital angular momentum direction. 
The crude estimates of this simple scenario give the stochastic variations, equations (\ref{eq:ConvRAM}) and (\ref{eq:ConvTAM}), to be several tens of percent of the constant component of the specific angular momentum (equation \ref{eq:ConstantAM}); this might account for our findings of the best angle distribution to fit the observed cumulative distribution function.  

 The stochastic variations that the convection motion introduces change on a time scale of the convective turnover time, $\simeq \tau_{\rm conv} \equiv l_c/v_{\rm conv}$. In AGB stars, the convection velocity is a large fraction of the sound speed, which is a large fraction of the Keplerian speed on the surface of the AGB star, $v_{\rm Kep}$. 
Therefore, the stochastic variation occurs on a time scale similar to the orbital period of a test particle on the surface of the AGB star or shorter, i.e., $ \tau_{\rm var} \simeq \tau_{\rm conv} \simeq 0.3 R_{\rm G}/v_{\rm conv} \approx R_{\rm G}/v_{\rm Kep} <  2 \pi R_{\rm G}/v_{\rm Kep} =\tau_{\rm Kep}$. The conclusion is that the convective motion can cause changes in the angular momentum axis in consecutive periastron passages, with a time-lapse of months to years. At the PN phase that occurs hundreds to thousands of years later, the jets will be observed to be coeval, like in the `starfish PNe', e.g., M1-37 (e.g., \citealt{Sahai2000}). The stochastic convective motion will also cause angular momentum fluctuation in cases where two or more jet-launching episodes are many years (many orbital periods) apart, as in KjPn8.

 \cite{Hsiaetal2014} suggest that the lobes result from the interactions between later-developed fast winds and previously ejected asymptotic giant branch winds. However, they do not specify the cause of the change in the directions of the symmetry axes. \cite{Steffenetal2013} simulate the interaction of a fast wind with a filamentary AGB shell. The filamentary structure is a stochastic one that is not point-symmetric. Namely, there are no pairs of opposite structures. They attribute the filamentary structure to instability or convective motion but not jets. Indeed, the multipolar structures they obtain are not point-symmetric. They do not obtain the structures of most of the PNe we study here, which are point-symmetric.

Future studies should explore the properties of the scenario we propose here and examine other scenarios, such as accretion at the CEE's termination or the star's tidal destruction at the core.

\section*{Acknowledgements}

We thank Gunter Cibis, Martin Guerrero, and an anonymous referee for their valuable comments. We heavily used the Planetary Nebula Image Catalogue (PNIC) of Bruce Balick \citep{Balick2006} \url{https://faculty.washington.edu/balick/PNIC/}.
A grant from the Pazy Foundation supported this research.


\begin{thebibliography}{}
\expandafter\ifx\csname natexlab\endcsname\relax\def\natexlab#1{#1}\fi
\providecommand{\url}[1]{\href{#1}{#1}}
\providecommand{\dodoi}[1]{doi:~\href{http://doi.org/#1}{\nolinkurl{#1}}}
\providecommand{\doeprint}[1]{\href{http://ascl.net/#1}{\nolinkurl{http://ascl.net/#1}}}
\providecommand{\doarXiv}[1]{\href{https://arxiv.org/abs/#1}{\nolinkurl{https://arxiv.org/abs/#1}}}

\bibitem[{{Ablimit}(2024)}]{Ablimit2024}
{Ablimit}, I. 2024, arXiv e-prints, arXiv:2407.03985, \dodoi{10.48550/arXiv.2407.03985}

\bibitem[{{Akashi} \& {Soker}(2018)}]{AkashiSoker2018}
{Akashi}, M., \& {Soker}, N. 2018, \mnras, 481, 2754, \dodoi{10.1093/mnras/sty2479}

\bibitem[{{Akras} \& {Gon{\c{c}}alves}(2016)}]{AkrasGonclves2016}
{Akras}, S., \& {Gon{\c{c}}alves}, D.~R. 2016, \mnras, 455, 930, \dodoi{10.1093/mnras/stv2139}

\bibitem[{{Baan} {et~al.}(2021){Baan}, {Imai}, \& {Orosz}}]{Baanetal2021}
{Baan}, W.~A., {Imai}, H., \& {Orosz}, G. 2021, Research in Astronomy and Astrophysics, 21, 275, \dodoi{10.1088/1674-4527/21/11/275}

\bibitem[{{Balick}(2006)}]{Balick2006}
{Balick}, B. 2006, {Planetary Nebula Image Catalogue: HST data}, HST Proposal ID 10933. Cycle 15

\bibitem[{{Balick} {et~al.}(2020){Balick}, {Frank}, \& {Liu}}]{Balicketal2020}
{Balick}, B., {Frank}, A., \& {Liu}, B. 2020, \apj, 889, 13, \dodoi{10.3847/1538-4357/ab5651}

\bibitem[{{Bandyopadhyay} {et~al.}(2023){Bandyopadhyay}, {Das}, {Parthasarathy}, \& {Kar}}]{Bandyopadhyayetal2023}
{Bandyopadhyay}, R., {Das}, R., {Parthasarathy}, M., \& {Kar}, S. 2023, \mnras, 524, 1547, \dodoi{10.1093/mnras/stad1897}

\bibitem[{{Boffin} \& {Jones}(2019)}]{BoffinJones2019}
{Boffin}, H. M.~J., \& {Jones}, D. 2019, {The Importance of Binaries in the Formation and Evolution of Planetary Nebulae}, \dodoi{10.1007/978-3-030-25059-1}

\bibitem[{{Boffin} {et~al.}(2012){Boffin}, {Miszalski}, {Rauch}, {Jones}, {Corradi}, {Napiwotzki}, {Day-Jones}, \& {K{\"o}ppen}}]{Boffinetal2012}
{Boffin}, H. M.~J., {Miszalski}, B., {Rauch}, T., {et~al.} 2012, Science, 338, 773, \dodoi{10.1126/science.1225386}

\bibitem[{{Chen} {et~al.}(2025){Chen}, {Fang}, {Chen}, \& {Liu}}]{ChenPetal2025}
{Chen}, P., {Fang}, X., {Chen}, X., \& {Liu}, J. 2025, arXiv e-prints, arXiv:2501.06056.
\newblock \doarXiv{2501.06056}

\bibitem[{{Chong} {et~al.}(2012){Chong}, {Kwok}, {Imai}, {Tafoya}, \& {Chibueze}}]{Chongetal2012}
{Chong}, S.~N., {Kwok}, S., {Imai}, H., {Tafoya}, D., \& {Chibueze}, J. 2012, \apj, 760, 115, \dodoi{10.1088/0004-637X/760/2/115}

\bibitem[{{Clairmont} {et~al.}(2022){Clairmont}, {Steffen}, \& {Koning}}]{Clairmontetal2022}
{Clairmont}, R., {Steffen}, W., \& {Koning}, N. 2022, \mnras, 516, 2711, \dodoi{10.1093/mnras/stac2375}

\bibitem[{{Clark} {et~al.}(2013){Clark}, {L{\'o}pez}, {Steffen}, \& {Richer}}]{Clarketal2013}
{Clark}, D.~M., {L{\'o}pez}, J.~A., {Steffen}, W., \& {Richer}, M.~G. 2013, \aj, 145, 57, \dodoi{10.1088/0004-6256/145/3/57}

\bibitem[{{Danehkar}(2022)}]{Danehkar2022}
{Danehkar}, A. 2022, \apjs, 260, 14, \dodoi{10.3847/1538-4365/ac5cca}

\bibitem[{{Derlopa} {et~al.}(2024){Derlopa}, {Akras}, {Amram}, {Boumis}, {Chiotellis}, \& {de Oliveira}}]{Derlopaetal2024}
{Derlopa}, S., {Akras}, S., {Amram}, P., {et~al.} 2024, \mnras, 530, 3327, \dodoi{10.1093/mnras/stae1013}

\bibitem[{{Estrella-Trujillo} {et~al.}(2019){Estrella-Trujillo}, {Hern{\'a}ndez-Mart{\'\i}nez}, {Vel{\'a}zquez}, {Esquivel}, \& {Raga}}]{EstrellaTrujilloetal2019}
{Estrella-Trujillo}, D., {Hern{\'a}ndez-Mart{\'\i}nez}, L., {Vel{\'a}zquez}, P.~F., {Esquivel}, A., \& {Raga}, A.~C. 2019, \apj, 876, 29, \dodoi{10.3847/1538-4357/ab12e1}

\bibitem[{{Gold} {et~al.}(2024){Gold}, {Schmidt}, \& {Ziurys}}]{Goldetal2024}
{Gold}, K.~R., {Schmidt}, D.~R., \& {Ziurys}, L.~M. 2024, \apj, 976, 196, \dodoi{10.3847/1538-4357/ad83be}

\bibitem[{{G{\'o}mez-Gonz{\'a}lez} {et~al.}(2020){G{\'o}mez-Gonz{\'a}lez}, {Toal{\'a}}, {Guerrero}, {Todt}, {Sabin}, {Ramos-Larios}, \& {Mayya}}]{GomezGonzalezetal2020}
{G{\'o}mez-Gonz{\'a}lez}, V.~M.~A., {Toal{\'a}}, J.~A., {Guerrero}, M.~A., {et~al.} 2020, \mnras, 496, 959, \dodoi{10.1093/mnras/staa1542}

\bibitem[{{G{\'o}mez-Mu{\~n}oz} {et~al.}(2023){G{\'o}mez-Mu{\~n}oz}, {V{\'a}zquez}, {Sabin}, {Olgu{\'\i}n}, {Guill{\'e}n}, {Zavala}, \& {Michel}}]{GomezMunozetal2023}
{G{\'o}mez-Mu{\~n}oz}, M.~A., {V{\'a}zquez}, R., {Sabin}, L., {et~al.} 2023, \aap, 676, A101, \dodoi{10.1051/0004-6361/202346455}

\bibitem[{{Guerrero} {et~al.}(2021){Guerrero}, {Cazzoli}, {Rechy-Garc{\'\i}a}, {Ramos-Larios}, {Montoro-Molina}, {G{\'o}mez-Gonz{\'a}lez}, {Toal{\'a}}, \& {Fang}}]{Guerreoetal2021}
{Guerrero}, M.~A., {Cazzoli}, S., {Rechy-Garc{\'\i}a}, J.~S., {et~al.} 2021, \apj, 909, 44, \dodoi{10.3847/1538-4357/abe2aa}

\bibitem[{{Guerrero} \& {Manchado}(1998)}]{Guerreroetal1998}
{Guerrero}, M.~A., \& {Manchado}, A. 1998, \apj, 508, 262, \dodoi{10.1086/306407}

\bibitem[{{Guerrero} {et~al.}(2013){Guerrero}, {Miranda}, {Ramos-Larios}, \& {V{\'a}zquez}}]{Guerreroetal2013}
{Guerrero}, M.~A., {Miranda}, L.~F., {Ramos-Larios}, G., \& {V{\'a}zquez}, R. 2013, \aap, 551, A53, \dodoi{10.1051/0004-6361/201220592}

\bibitem[{{Guerrero} {et~al.}(2020){Guerrero}, {Suzett Rechy-Garc{\'\i}a}, \& {Ortiz}}]{Guerreroetal2020}
{Guerrero}, M.~A., {Suzett Rechy-Garc{\'\i}a}, J., \& {Ortiz}, R. 2020, \apj, 890, 50, \dodoi{10.3847/1538-4357/ab61fa}

\bibitem[{{Guerrero} {et~al.}(2008){Guerrero}, {Miranda}, {Riera}, {Vel{\'a}zquez}, {Olgu{\'\i}n}, {V{\'a}zquez}, {Chu}, {Raga}, \& {Ben{\'\i}tez}}]{Guerrero2008}
{Guerrero}, M.~A., {Miranda}, L.~F., {Riera}, A., {et~al.} 2008, \apj, 683, 272, \dodoi{10.1086/588632}

\bibitem[{{Guill{\'e}n} {et~al.}(2013){Guill{\'e}n}, {V{\'a}zquez}, {Miranda}, {Zavala}, {Contreras}, {Ayala}, \& {Ortiz-Ambriz}}]{Guillenetal2013}
{Guill{\'e}n}, P.~F., {V{\'a}zquez}, R., {Miranda}, L.~F., {et~al.} 2013, \mnras, 432, 2676, \dodoi{10.1093/mnras/stt612}

\bibitem[{{Harman} {et~al.}(2004){Harman}, {Bryce}, {L{\'o}pez}, {Meaburn}, \& {Holloway}}]{Harmanetal2004}
{Harman}, D.~J., {Bryce}, M., {L{\'o}pez}, J.~A., {Meaburn}, J., \& {Holloway}, A.~J. 2004, \mnras, 348, 1047, \dodoi{10.1111/j.1365-2966.2004.07427.x}

\bibitem[{{Hillwig} {et~al.}(2017){Hillwig}, {Frew}, {Reindl}, {Rotter}, {Webb}, \& {Margheim}}]{Hillwigetal2017}
{Hillwig}, T.~C., {Frew}, D.~J., {Reindl}, N., {et~al.} 2017, \aj, 153, 24, \dodoi{10.3847/1538-3881/153/1/24}

\bibitem[{{Hrivnak} {et~al.}(1999){Hrivnak}, {Kwok}, \& {Su}}]{Hrivnaketal1999}
{Hrivnak}, B.~J., {Kwok}, S., \& {Su}, K. Y.~L. 1999, \apj, 524, 849, \dodoi{10.1086/307822}

\bibitem[{{Hsia} {et~al.}(2014){Hsia}, {Chau}, {Zhang}, \& {Kwok}}]{Hsiaetal2014}
{Hsia}, C.-H., {Chau}, W., {Zhang}, Y., \& {Kwok}, S. 2014, \apj, 787, 25, \dodoi{10.1088/0004-637X/787/1/25}

\bibitem[{{Huang} {et~al.}(2016){Huang}, {Lee}, {Moraghan}, \& {Smith}}]{Huangetal2016}
{Huang}, P.-S., {Lee}, C.-F., {Moraghan}, A., \& {Smith}, M. 2016, \apj, 820, 134, \dodoi{10.3847/0004-637X/820/2/134}

\bibitem[{{Jones}(2024)}]{Jones2025}
{Jones}, D. 2024, arXiv e-prints, arXiv:2411.06831, \dodoi{10.48550/arXiv.2411.06831}

\bibitem[{{Jones} \& {Boffin}(2017)}]{JonesBoffin2017}
{Jones}, D., \& {Boffin}, H. M.~J. 2017, Nature Astronomy, 1, 0117, \dodoi{10.1038/s41550-017-0117}

\bibitem[{{Jones} {et~al.}(2023){Jones}, {Hillwig}, \& {Reindl}}]{Jonesetal2023}
{Jones}, D., {Hillwig}, T.~C., \& {Reindl}, N. 2023, in Highlights on Spanish Astrophysics XI, ed. M.~{Manteiga}, L.~{Bellot}, P.~{Benavidez}, A.~{de Lorenzo-C{\'a}ceres}, M.~A. {Fuente}, M.~J. {Mart{\'\i}nez}, M.~{V{\'a}zquez Acosta}, \& C.~{Dafonte}, 216, \dodoi{10.48550/arXiv.2304.06355}

\bibitem[{{Jones} {et~al.}(2020){Jones}, {Boffin}, {Hibbert}, {Steinmetz}, {Wesson}, {Hillwig}, {Sowicka}, {Corradi}, {Garc{\'\i}a-Rojas}, {Rodr{\'\i}guez-Gil}, \& {Munday}}]{Jonesetal2020}
{Jones}, D., {Boffin}, H.~M.~J., {Hibbert}, J., {et~al.} 2020, \aap, 642, A108, \dodoi{10.1051/0004-6361/202038778}

\bibitem[{{Jones} {et~al.}(2022){Jones}, {Munday}, {Corradi}, {Rodr{\'\i}guez-Gil}, {Boffin}, {Zak}, {Sowicka}, {Parsons}, {Dhillon}, {Littlefair}, {Marsh}, {Reindl}, \& {Garc{\'\i}a-Rojas}}]{Jonesetal2022}
{Jones}, D., {Munday}, J., {Corradi}, R. L.~M., {et~al.} 2022, \mnras, 510, 3102, \dodoi{10.1093/mnras/stab3736}

\bibitem[{{Kwok}(2024)}]{Kwok2024}
{Kwok}, S. 2024, Galaxies, 12, 39, \dodoi{10.3390/galaxies12040039}

\bibitem[{{Kwok} {et~al.}(2010){Kwok}, {Chong}, {Hsia}, {Zhang}, \& {Koning}}]{Kwoketal2010}
{Kwok}, S., {Chong}, S.-N., {Hsia}, C.-H., {Zhang}, Y., \& {Koning}, N. 2010, \apj, 708, 93, \dodoi{10.1088/0004-637X/708/1/93}

\bibitem[{{L{\'o}pez} {et~al.}(1998){L{\'o}pez}, {Meaburn}, {Bryce}, \& {Holloway}}]{Lopezetal1998}
{L{\'o}pez}, J.~A., {Meaburn}, J., {Bryce}, M., \& {Holloway}, A.~J. 1998, \apj, 493, 803, \dodoi{10.1086/305155}

\bibitem[{{L{\'o}pez} {et~al.}(2000){L{\'o}pez}, {Meaburn}, {Rodr{\'\i}guez}, {V{\'a}zquez}, {Steffen}, \& {Bryce}}]{Lopezetal2000}
{L{\'o}pez}, J.~A., {Meaburn}, J., {Rodr{\'\i}guez}, L.~F., {et~al.} 2000, \apj, 538, 233, \dodoi{10.1086/309122}

\bibitem[{{Manchado} {et~al.}(1996{\natexlab{a}}){Manchado}, {Guerrero}, {Stanghellini}, \& {Serra-Ricart}}]{Manchadoetal1996}
{Manchado}, A., {Guerrero}, M.~A., {Stanghellini}, L., \& {Serra-Ricart}, M. 1996{\natexlab{a}}, {The IAC morphological catalog of northern Galactic planetary nebulae}

\bibitem[{{Manchado} {et~al.}(1996{\natexlab{b}}){Manchado}, {Stanghellini}, \& {Guerrero}}]{Manchadoetal1996b}
{Manchado}, A., {Stanghellini}, L., \& {Guerrero}, M.~A. 1996{\natexlab{b}}, \apjl, 466, L95, \dodoi{10.1086/310170}

\bibitem[{{Miranda} {et~al.}(2001){Miranda}, {Guerrero}, \& {Torrelles}}]{Mirandaetal2001}
{Miranda}, L.~F., {Guerrero}, M.~A., \& {Torrelles}, J.~M. 2001, \mnras, 322, 195, \dodoi{10.1046/j.1365-8711.2001.04145.x}

\bibitem[{{Miranda} {et~al.}(1998){Miranda}, {Torrelles}, {Guerrero}, {Aaquist}, \& {Eiroa}}]{Mirandaetal1998}
{Miranda}, L.~F., {Torrelles}, J.~M., {Guerrero}, M.~A., {Aaquist}, O.~B., \& {Eiroa}, C. 1998, \mnras, 298, 243, \dodoi{10.1046/j.1365-8711.1998.01611.x}

\bibitem[{{Miranda} {et~al.}(2024){Miranda}, {V{\'a}zquez}, {Olgu{\'\i}n}, {Guill{\'e}n}, \& {Mat{\'\i}as}}]{Mirandaetal2024}
{Miranda}, L.~F., {V{\'a}zquez}, R., {Olgu{\'\i}n}, L., {Guill{\'e}n}, P.~F., \& {Mat{\'\i}as}, J.~M. 2024, \aap, 687, A123, \dodoi{10.1051/0004-6361/202348173}

\bibitem[{{Moraga Baez} {et~al.}(2023){Moraga Baez}, {Kastner}, {Balick}, {Montez}, \& {Bublitz}}]{MoragaBaezetal2023}
{Moraga Baez}, P., {Kastner}, J.~H., {Balick}, B., {Montez}, R., \& {Bublitz}, J. 2023, \apj, 942, 15, \dodoi{10.3847/1538-4357/aca401}

\bibitem[{{Morris}(1987)}]{Morris1987}
{Morris}, M. 1987, \pasp, 99, 1115, \dodoi{10.1086/132089}

\bibitem[{{Rechy-Garc{\'\i}a} {et~al.}(2020){Rechy-Garc{\'\i}a}, {Guerrero}, {Duarte Puertas}, {Chu}, {Toal{\'a}}, \& {Miranda}}]{RechyGarciaetal2020}
{Rechy-Garc{\'\i}a}, J.~S., {Guerrero}, M.~A., {Duarte Puertas}, S., {et~al.} 2020, \mnras, 492, 1957, \dodoi{10.1093/mnras/stz3326}

\bibitem[{{Rechy-Garc{\'\i}a} {et~al.}(2019){Rechy-Garc{\'\i}a}, {Pe{\~n}a}, \& {Vel{\'a}zquez}}]{RechyGarciaetal2019}
{Rechy-Garc{\'\i}a}, J.~S., {Pe{\~n}a}, M., \& {Vel{\'a}zquez}, P.~F. 2019, \mnras, 482, 1163, \dodoi{10.1093/mnras/sty2758}

\bibitem[{{Rubio} {et~al.}(2015){Rubio}, {V{\'a}zquez}, {Ramos-Larios}, {Guerrero}, {Olgu{\'\i}n}, {Guill{\'e}n}, \& {Mata}}]{Rubioetal2015}
{Rubio}, G., {V{\'a}zquez}, R., {Ramos-Larios}, G., {et~al.} 2015, \mnras, 446, 1931, \dodoi{10.1093/mnras/stu2201}

\bibitem[{{Sabin} {et~al.}(2012){Sabin}, {V{\'a}zquez}, {L{\'o}pez}, {Garc{\'\i}a-D{\'\i}az}, \& {Ramos-Larios}}]{Sabinetal2012}
{Sabin}, L., {V{\'a}zquez}, R., {L{\'o}pez}, J.~A., {Garc{\'\i}a-D{\'\i}az}, M.~T., \& {Ramos-Larios}, G. 2012, \rmxaa, 48, 165, \dodoi{10.48550/arXiv.1203.1297}

\bibitem[{{Sahai}(2000)}]{Sahai2000}
{Sahai}, R. 2000, \apjl, 537, L43, \dodoi{10.1086/312748}

\bibitem[{{Sahai} {et~al.}(2005{\natexlab{a}}){Sahai}, {Le Mignant}, {S{\'a}nchez Contreras}, {Campbell}, \& {Chaffee}}]{Sahaietal2005}
{Sahai}, R., {Le Mignant}, D., {S{\'a}nchez Contreras}, C., {Campbell}, R.~D., \& {Chaffee}, F.~H. 2005{\natexlab{a}}, \apjl, 622, L53, \dodoi{10.1086/429586}

\bibitem[{{Sahai} {et~al.}(2007{\natexlab{a}}){Sahai}, {Morris}, {S{\'a}nchez Contreras}, \& {Claussen}}]{Sahaietal2007a}
{Sahai}, R., {Morris}, M., {S{\'a}nchez Contreras}, C., \& {Claussen}, M. 2007{\natexlab{a}}, \aj, 134, 2200, \dodoi{10.1086/522944}

\bibitem[{{Sahai} {et~al.}(2011){Sahai}, {Morris}, \& {Villar}}]{Sahaietal20112011}
{Sahai}, R., {Morris}, M.~R., \& {Villar}, G.~G. 2011, \aj, 141, 134, \dodoi{10.1088/0004-6256/141/4/134}

\bibitem[{{Sahai} {et~al.}(2005{\natexlab{b}}){Sahai}, {S{\'a}nchez Contreras}, \& {Morris}}]{Sahaietal2005Starfish}
{Sahai}, R., {S{\'a}nchez Contreras}, C., \& {Morris}, M. 2005{\natexlab{b}}, \apj, 620, 948, \dodoi{10.1086/426469}

\bibitem[{{Sahai} {et~al.}(2007{\natexlab{b}}){Sahai}, {S{\'a}nchez Contreras}, {Morris}, \& {Claussen}}]{Sahaietal2007b}
{Sahai}, R., {S{\'a}nchez Contreras}, C., {Morris}, M., \& {Claussen}, M. 2007{\natexlab{b}}, \apj, 658, 410, \dodoi{10.1086/511294}

\bibitem[{{Sahai} \& {Trauger}(1998)}]{SahaiTrauger1998}
{Sahai}, R., \& {Trauger}, J.~T. 1998, \aj, 116, 1357, \dodoi{10.1086/300504}

\bibitem[{{Sahai} {et~al.}(2017){Sahai}, {Vlemmings}, \& {Nyman}}]{Sahaietal2017}
{Sahai}, R., {Vlemmings}, W.~H.~T., \& {Nyman}, L.~{\r{A}}. 2017, \apj, 841, 110, \dodoi{10.3847/1538-4357/aa6d86}

\bibitem[{{Sahai} {et~al.}(2024){Sahai}, {Alcolea}, {Balick}, {Blackman}, {Bujarrabal}, {Castro-Carrizo}, {De Marco}, {Kastner}, {Kim}, {Lagadec}, {Lee}, {Sabin}, {Santander-Garcia}, {S{\'a}nchez Contreras}, {Tafoya}, {Ueta}, {Vlemmings}, \& {Zijlstra}}]{Sahaietal2024}
{Sahai}, R., {Alcolea}, J., {Balick}, B., {et~al.} 2024, arXiv e-prints, arXiv:2409.06038.
\newblock \doarXiv{2409.06038}

\bibitem[{{Schwarz} {et~al.}(1992){Schwarz}, {Corradi}, \& {Melnick}}]{Schwarzetal1992}
{Schwarz}, H.~E., {Corradi}, R.~L.~M., \& {Melnick}, J. 1992, \aaps, 96, 23

\bibitem[{{Soker}(1990)}]{Soker1990AJ}
{Soker}, N. 1990, \aj, 99, 1869, \dodoi{10.1086/115465}

\bibitem[{{Soker}(2023{\natexlab{a}})}]{Soker2025Bright}
---. 2023{\natexlab{a}}, arXiv e-prints, arXiv:2310.15785, \dodoi{10.48550/arXiv.2310.15785}

\bibitem[{{Soker}(2023{\natexlab{b}})}]{Soker2023IAUPN}
---. 2023{\natexlab{b}}, arXiv e-prints, arXiv:2310.15785, \dodoi{10.48550/arXiv.2310.15785}

\bibitem[{{Sowicka} {et~al.}(2017){Sowicka}, {Jones}, {Corradi}, {Wesson}, {Garc{\'\i}a-Rojas}, {Santander-Garc{\'\i}a}, {Boffin}, \& {Rodr{\'\i}guez-Gil}}]{Sowickaetal2017}
{Sowicka}, P., {Jones}, D., {Corradi}, R. L.~M., {et~al.} 2017, \mnras, 471, 3529, \dodoi{10.1093/mnras/stx1697}

\bibitem[{{Steffen} {et~al.}(2013){Steffen}, {Koning}, {Esquivel}, {Garc{\'\i}a-Segura}, {Garc{\'\i}a-D{\'\i}az}, {L{\'o}pez}, \& {Magnor}}]{Steffenetal2013}
{Steffen}, W., {Koning}, N., {Esquivel}, A., {et~al.} 2013, \mnras, 436, 470, \dodoi{10.1093/mnras/stt1583}

\bibitem[{{Tafoya} {et~al.}(2019){Tafoya}, {Orosz}, {Vlemmings}, {Sahai}, \& {P{\'e}rez-S{\'a}nchez}}]{Tafoyaetal2019}
{Tafoya}, D., {Orosz}, G., {Vlemmings}, W.~H.~T., {Sahai}, R., \& {P{\'e}rez-S{\'a}nchez}, A.~F. 2019, \aap, 629, A8, \dodoi{10.1051/0004-6361/201834632}

\bibitem[{{Trammell} \& {Goodrich}(2002)}]{Trammelletal2002}
{Trammell}, S.~R., \& {Goodrich}, R.~W. 2002, \apj, 579, 688, \dodoi{10.1086/342943}

\bibitem[{{Vel{\'a}zquez} {et~al.}(2012){Vel{\'a}zquez}, {Raga}, {Riera}, {Steffen}, {Esquivel}, {Cant{\'o}}, \& {Haro-Corzo}}]{Velazquez2012}
{Vel{\'a}zquez}, P.~F., {Raga}, A.~C., {Riera}, A., {et~al.} 2012, \mnras, 419, 3529, \dodoi{10.1111/j.1365-2966.2011.19991.x}

\bibitem[{{Wen} {et~al.}(2024){Wen}, {Wang}, {Hsia}, {Yeh}, {Liu}, {Liu}, \& {Kang}}]{Wenetal2024}
{Wen}, S., {Wang}, Y.-Z., {Hsia}, C.-H., {et~al.} 2024, \aap, 687, A99, \dodoi{10.1051/0004-6361/202449751}

\bibitem[{{Wen} {et~al.}(2023){Wen}, {Hsia}, {Kang}, {Chen}, \& {Luo}}]{Wenetal2023}
{Wen}, S.-B., {Hsia}, C.-H., {Kang}, X.-X., {Chen}, R., \& {Luo}, T. 2023, Research in Astronomy and Astrophysics, 23, 035018, \dodoi{10.1088/1674-4527/acbe95}

\end{thebibliography}

\end{document}